\documentclass[12pt]{article}
\usepackage{amsmath,epsfig,amssymb}
\usepackage{latexsym}
\usepackage{graphicx}
\usepackage{subfigure}
\usepackage{overpic}

\begin{document}

\title{\vbox{
\baselineskip 14pt
\hfill \hbox{\normalsize KUNS-2316}\\
\hfill \hbox{\normalsize YITP-10-97} } \vskip 2cm
\bf Inflation, moduli (de)stabilization \\ and supersymmetry breaking \\ 
 \vskip 0.5cm
}
\author{
Tatsuo~Kobayashi$^{1,}$\footnote{
email: kobayash@gauge.scphys.kyoto-u.ac.jp} \ 
  \ and \
Manabu~Sakai$^{2,}$\footnote{email: msakai@yukawa.kyoto-u.ac.jp}
\\*[20pt]
$^1${\it \normalsize
Department of Physics, Kyoto University,
Kyoto 606-8502, Japan} \\
$^2${\it \normalsize 
Yukawa Institute for Theoretical Physics, Kyoto University, 
Kyoto 606-8502, Japan}
}

\date{}

\maketitle
\thispagestyle{empty}

\begin{abstract}
We study the cosmological inflation from the 
viewpoint of the moduli stabilization.
We study the scenario that the superpotential 
has a large value during the inflation era enough to 
stabilize moduli, but it is small in 
the true vacuum.
This scenario is discussed by using a simple model, 
one type of hybrid models.
\end{abstract}

\newpage

\section{Introduction}

Moduli fields play an important role in superstring theory.
Couplings in 4D low-energy effective field theory are given 
as functions of vacuum expectation values (VEVs) of moduli fields.
Thus, we need a stabilization mechanism of moduli VEVs.  
Indeed, moduli stabilization is one of important issues 
in string phenomenology and cosmology.
The moduli potential has a small bump, which is related to 
the gravitino mass $m_{3/2}$, in many models of 
moduli stabilization, e.g. the Kachru-Kallosh-Linde-Trivedi
(KKLT) potential \cite{Kachru:2003aw} and 
the racetrack potential \cite{racetrack}.

When we consider the low-energy supersymmetry (SUSY) breaking scenario, 
the height of the above bump is quite low.
That may lead to some problems.
For example, we need the positive energy to derive 
inflational expansion of the Universe.
If such a inflation energy is larger than this small bump 
of the moduli potential, 
the moduli would be destabilized and run away to 
infinity.
Thus, we may have a constraint between the gravitino mass $m_{3/2}$ and 
the Hubble constant $H_{inf}$ 
during inflation, e.g. $m_{3/2} \geq H_{inf}$ in a simple model 
\cite{Kallosh:2004yh}.

One way out is to use non-perturbative superpotential 
with positive exponents \cite{Abe:2005rx,Abe:2008xu,Badziak:2008gv}, 
although the KKLT superpotential and the usual racetrack
superpotential include non-perturbative terms with negative 
exponents.
However, the positive exponents are possible and 
have significant effects in the moduli stabilization.
That is, the moduli potential with positive exponents could have
a quite high barrier, which is independent of 
the gravitino mass.

In this paper, we study another scenario to stabilize 
the moduli during the inflation era and to lead to low-energy SUSY.
We consider the inflation scenario that  
the inflaton field $\phi$, which is different from moduli, 
drives the inflation dominantly.
We assume that the would-be inflaton $\phi$ induces a large value of 
superpotential $\langle W(\phi) \rangle_{inf}$ during the 
inflation era.
A similar idea has been studied in \cite{He:2010uk}.
At any rate, such a large value of $\langle W(\phi) \rangle_{inf}$ could also 
induce a large mass of the modulus during the inflation era, 
and the modulus mass during the inflation would be of ${\cal
  O}(10)\times \langle W(\phi) \rangle_{inf}$, 
e.g.  for the KKLT superpotential.
Here we use the unit that $M_{Pl}=1$, where $M_{Pl}$ denotes 
the Planck scale.
If such a mass is larger than the Hubble parameter $H_{inf}$ 
during the inflation era, the modulus is not be destabilized.\footnote{
This behavior is similar to the F-term uplifting scenario 
\cite{Saltman:2004sn,Dudas:2006gr,Abe:2006xp,Kallosh:2006dv}.}
After the inflation ends, we assume that a small value of 
the gravitino mass $m_{3/2}$, i.e. $m_{3/2} \ll \langle W(\phi)
\rangle_{inf}$,  is realized at the potential minimum.
At such a potential minimum, the modulus has a large mass 
compared with the gravitino mass $m_{3/2}$ such as 
${\cal O}(10)\times  m_{3/2}$ or more.
Then, the modulus is stabilized at the true minimum.
That is the realization of the F-term uplifting scenario 
at the potential minimum \cite{Saltman:2004sn,Dudas:2006gr,Abe:2006xp,Kallosh:2006dv}.

To illustrate our scenario, we use the inflation model studied in 
Ref.~\cite{Nakai:2010km}, which is a kind of 
hybrid inflation models \cite{Dvali:1994ms,Dimopoulos:1997fv}.
Its inflaton superpotential $W(\phi)$ is 
the Intrilligator-Seiberg-Shih (ISS) type of superpotential 
\cite{Intriligator:2006dd} with a deformation proposed 
in Ref.~\cite{Kitano:2006xg}.
(See also for the inflation model with the ISS superpotential 
\cite{Craig:2008tv}.) 
With such a inflation superpotential $W(\phi)$, 
we study the modulus behavior during the inflation era 
for the KKLT superpotential and the racetrack superpotential 
including the case with a positive exponent.
Indeed, similar superpotential models have been studied 
for inflation models \cite{hybrid-moduli}.
However, our viewpoint differs from those.

This paper is organized as follows.
In section 2, we explain our scenario without explicit 
inflation superpotential.
In section 3, we study our scenario by using an illustrating model.
Section 4 is devoted to conclusion and discussion.

\section{Scenario}

Here, we outline our scenario.
We assume a superpotential $W(\phi)$ for 
the would-be inflaton $\phi$, which induces 
a large Hubble parameter $H_{inf}$ in the inflation era 
(without taking into account the degree of freedom of the moduli).
Then, we study the degree of freedom of the moduli.
For simplicity, we consider a single modulus, $T$, with 
its K\"ahler potential,
\begin{eqnarray}
-3 \ln (T + T^*).
\end{eqnarray}
Then, the total K\"ahler potential is written by
\begin{eqnarray}
K= -3 \ln (T + T^*) + K(\phi,\phi^*),
\end{eqnarray}
where the second term denotes the K\"ahler potential of $\phi$ 
and a simple model may correspond to the canonical form, 
$K(\phi,\phi^*) =|\phi|^2$.
The modulus $T$ cannot be stabilized without non-perturbative terms.
Thus, we introduce a certain non-perturbative term.
First, we consider the KKLT-type superpotential.
That is, the total superpotential is written by
\begin{eqnarray}\label{eq:KKLT-W}
W=W(\phi) - Ae^{-aT},
\end{eqnarray}
with $a >0$.
Here, we assume  $a={\cal O}(10)$, which is a natural 
value when this non-perturbative term is induced e.g. 
by gaugino condensation.
Then, the total scalar potential is written as
\begin{eqnarray}
V=e^K[|D_IW|^2(K^{I\bar I}) -3 |W|^2],
\end{eqnarray}
where $D_IW = K_IW + W_I$ and $K^{I\bar I}$ denotes 
the inverse of the K\"ahler metric $K_{I\bar I}$.

For the moment, we neglect the dynamics of $\phi$.
That is, we replace $\phi$ by a fixed value and 
neglect $D_\phi W$, i.e. $D_\phi W=0$ in the scalar potential.
In such a case, the modulus $T$ can be stabilized 
at $D_TW=0$, i.e.,
\begin{eqnarray}\label{eq:DTW=0}
K_T(W(\phi) - Ae^{-aT})+aAe^{-aT}=0,
\end{eqnarray}
where $W(\phi)$ is considered as a constant.
When $|aT| \gg 1$, this equation becomes 
\begin{eqnarray}\label{eq:DTW=0-2}
aT \approx  -log(W(\phi))-log(a(T+T^*)/3),
\end{eqnarray}
and we obtain $|W(\phi)| \gg |Ae^{-aT}|$, 
that is, the term  $W(\phi)$ is dominant in the superpotential.
To realize $|aT| \gg 1$, the superpotential $W(\phi)$ must be 
small such as $|W(\phi)| \ll 1$.
At this point, the modulus has a mass 
\begin{eqnarray}\label{eq:modulus-mass}
m_T=e^{k/2}(K^{T T^*})W_{TT}=a^2Ae^{-aT}e^{k/2}(K^{T 
  T^*})\approx  
\frac{a}{\sqrt{T + T^*}} W(\phi).
\end{eqnarray}

Now, let us take into account the dynamics of $\phi$, 
that is, non-vanishing $D_\phi W$.
We assume that through non-vanishing $D_\phi W$ 
the inflaton field $\phi$ induces 
a large positive energy to drive the inflation.
That is, a large Hubble parameter $H_{inf}$ is induced.
Such a large Hubble parameter may destabilize the modulus 
and the modulus may run away to infinity.
Once the modulus is destabilized in such a way, 
the successful inflation cannot be realized.
However, the modulus has the mass $m_T \approx a W(\phi)/\sqrt{T +
  T^*}$ 
at the above point (\ref{eq:DTW=0-2}).
Thus, the modulus would be stabilized still almost at 
the above point (\ref{eq:DTW=0-2}), when such a modulus mass is large enough 
compared with the Hubble parameter, i.e. 
$m_T \gg H_{inf}$. 
Then, the successful inflation could be realized.
More precisely speaking, the value of $T$ is changing during inflation 
as the would-be inflaton $\phi$ is moving along a slow-roll direction.
Hence, the true inflaton is a linear combination of $T$ and $\phi$.
However, the $\phi$ direction is dominant in the true inflaton 
because of the large mass $m_T$ of $T$.
The keypoint to realize our scenario is the 
inflation model with a large value of $W(\phi)$ 
such that $a W(\phi)/\sqrt{T +
  T^*} \gg H_{inf}$ during the inflation era.
After the inflation, the system approaches toward 
the true minimum.
We assume that at the true minimum a small gravitino mass, 
$m_{3/2} \ll \langle W(\phi) \rangle_{inf}$, 
and low-energy SUSY breaking are realized.

Figures 1 and 2 show what we expect in our scenario. 
That is, the would-be inflaton $\phi$ has a flat potential 
with a large energy corresponding to $H_{inf}$ 
in a certain region, and that would drive the inflation.
In the same region, the value of superpotential $W_{inf}$ is also 
large and it induces a large mass along the modulus direction 
like $m_T = {\cal O}(10) W_{inf}$.
After the inflation ends, the system approaches toward 
the true vacuum, which corresponds to $\phi = 0$ 
in these figures.
At that point, we would have a gravitino mass smaller than 
$W_{inf}$ and $H_{inf}$.


\begin{figure}
\begin{center}
\begin{overpic}[width=8cm,clip]{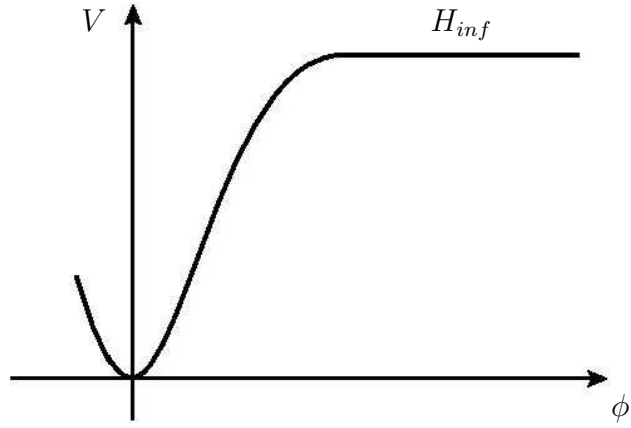}
\put(12,65){$V$}
\put(100,1){$\phi$}
\put(70,65){$H_{inf}$}
\end{overpic}
\caption{The scalar potential}
\end{center}
\end{figure}

\begin{figure}
\begin{center}
\begin{overpic}[width=8cm,clip]{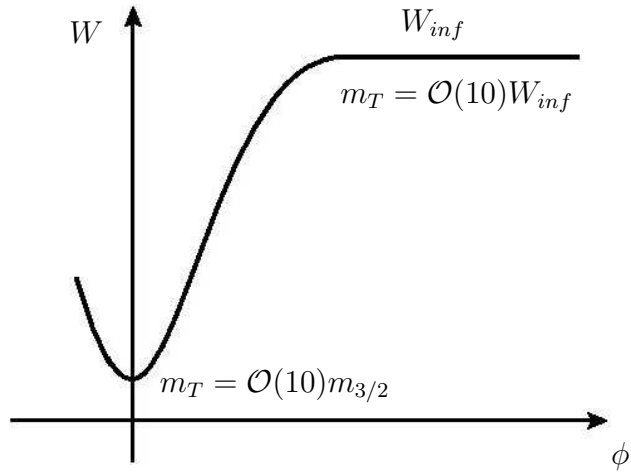}
\put(10,70){$W$}
\put(100,0){$\phi$}
\put(65,72){$W_{inf}$}
\put(55,60){$m_T=\mathcal{O}(10)W_{inf}$}
\put(25,12){$m_T=\mathcal{O}(10)m_{3/2}$}
\end{overpic}
\caption{The superpotential}
\end{center}
\end{figure}

Furthermore, we comment on the superpotential 
with a positive exponent, e.g. 
\begin{eqnarray}\label{eq:KKLT-W-2}
W=W(\phi) + Ae^{aT},
\end{eqnarray}
with $a >0$.
Such a non-perturbative superpotential may occur 
by gaugino condensation, where its gauge kinetic 
function $f$ is a linear combination between $S$ and $T$,
\begin{eqnarray}
f = mS - wT,
\end{eqnarray}
with $m,w >0$, and the field $S$ is stabilized already by its heavy
mass.
(See for details \cite{Abe:2005rx,Abe:2008xu}.)
In this case, we may have $A \ll 1$ in Eq.~(\ref{eq:KKLT-W-2}).
With this superpotential, the modulus $T$ has the mass 
$m_T=aW(\phi)/\sqrt{T +
  T^*}$ around $D_T W =0$ like Eq.~(\ref{eq:modulus-mass}).
Then, when $m_T \gg H_{inf}$, the successful inflation would be 
realized as the case with the superpotential (\ref{eq:KKLT-W-2}).
The modulus mass $m_T$ is the same between the superpotentials 
with the negative exponent (\ref{eq:KKLT-W}) and 
positive exponent (\ref{eq:KKLT-W-2}).
In both superpotentials, the successful inflation would occur 
similarly.
However, the forms of those scalar potentials are different from 
each other as we will show later by an explicit form of $W(\phi)$.
We can also consider the following superpotential 
\begin{eqnarray}\label{eq:racetrack-W-2}
W=W(\phi) + Ae^{-aT} +Be^{bT},
\end{eqnarray}
with $a,b >0$, where we may have $B \ll 1$ similar to 
$A$  in Eq.~(\ref{eq:KKLT-W-2}).

\section{A simple model}

Here, we illustrate our scenario by using a simple model.
As such an illustrating model for the inflation superpotential 
$W(\phi)$, we use the deformed ISS inflation \cite{Nakai:2010km}.
(See also \cite{Craig:2008tv}.)

\subsection{Deformed ISS inflation}

First, we briefly review on the (deformed) ISS inflation model.
This model is the supersymmetric $SU(N)$ gauge theory with 
$N_f$ flavors of fundamental and anti-fundamental matter fields, 
$q_i^c$ and $\bar{q}_j^c$ and singlets, $M_{ij}$, where 
$i,j = 1 \ldots N_f, c = 1 \ldots N$.
Their superpotential is written by 
\begin{eqnarray}
W_{ISS} = w_0+ h q_i^c M_{i j} \bar{q}_j^c - m^2 M_{I I} - \mu^2 M_{a a},
\end{eqnarray}
where $I = 1 \ldots N, a = N+1 \ldots N_f$ and $w_0$ is a constant.
We assume
\begin{eqnarray}\label{eq:mu-m}
 \ w_0 \sim \mu^2 \ll m^2 \ll 1 ,
\end{eqnarray}
and these fields have canonical K\"ahler potential:\footnote{
If higher order terms are large, this model would not satisfy 
the slow-roll conditions.
It is studied that the higher order terms are small enough 
in the (deformed) ISS inflation \cite{Craig:2008tv,Nakai:2010km}.
At any rate, we are using the deformed ISS inflation as 
an illustrating model for our scenario.}
\begin{eqnarray}
K_{ISS} = |M_{ij}|^2 + |q_i^c|^2 + |\bar{q}_i^c|^2.
\end{eqnarray}

Now, let us decompose fields as follow:
\begin{equation}
M_{i j} =
\begin{pmatrix}
Y_{I J} & Z_{I a} \\
\bar{Z}_{a I} & \Phi_{ab}
\end{pmatrix}
, \quad
q_i^c =
\begin{pmatrix}
\chi_{I}^c & \rho_{a}^c
\end{pmatrix}
, \quad
\bar{q}_i^c =
\begin{pmatrix}
\bar{\chi}_{I}^c \\
\bar{\rho}_{a}^c
\end{pmatrix}.
\end{equation}
The SUSY conditions, $D_I W =0$, for all fields cannot be 
satisfied by the rank condition.
The minimum corresponds to the point, which satisfies 
the conditions, $D_{\chi} W=D_\rho W=D_Z W=D_Y W=0$, 
that is, 
\begin{eqnarray}
\rho = \bar{\rho} = Z = \bar{Z} = 0, \ hY = -W, \ \chi = \bar{\chi} = \sqrt{m^2/h +|Y|^2},
\end{eqnarray}
but does not satisfy $D_\Phi W=0$.
That is the so-called ISS vacuum.
Thus, at such a vacuum the SUSY is broken by 
F-terms of $\Phi_{ab}$.
We may tune the constant $w_0$ such that the almost vanishing 
vacuum energy, $V_0 \sim 0$, at this vacuum is realized 
after including the modulus potential.

Here, we study the inflation in this potential.
Indeed, in this model, the would-be inflaton $\phi$ 
corresponds to the diagonal direction of $Y_{IJ}$, 
i.e., $Y_{I J} = \frac{1}{\sqrt N}\phi \delta_{I J} $ 
\cite{Nakai:2010km}.
In the global SUSY limit, the relevant term of scalar potential 
can be written as 
\begin{equation}
V_{tree} \simeq \left| m^2 \sqrt{N} - \frac{h}{\sqrt{N}} 
\chi \bar{\chi} \right|^2 + \frac{h^2}{N} |\phi|^2 
( |\chi|^2 + |\bar{\chi}|^2 ) ,
\end{equation}
at the tree-level, where we have also neglected $\mu$.
{}From now one, we concentrate ourselves on only the real part of 
$\phi$.
Along $\chi = \bar \chi =0$, the field $\phi$ has the almost 
flat potential.
When $\phi > \phi_c \equiv \sqrt{\frac{N}{h}}m$, 
the fields $\chi$ and $\bar \chi$ become massive.
Then, $\phi$ field comes to have the following potential 
by integrating out $\chi,\bar{\chi}$:
\begin{equation}
\begin{split}
V_{eff}(\phi)
&= N m^4 + \frac{N^2}{32\pi^2 } \Biggl[ 2 h^2 m^4 \log{\left( \frac{h^2 \phi^2}{NM_{\ast}^2} \right)} \\
&\quad+ \left( \frac{h^2}{N} \phi^2 + hm^2 \right)^2 \log{\left( 1 + \frac{Nm^2}{h\phi^2} \right)} + \left( \frac{h^2}{N} \phi^2 - hm^2 \right)^2 \log{\left( 1 - \frac{Nm^2}{h\phi^2} \right)} \Biggr],
\end{split}
\end{equation}
where $M_{\ast}$ denotes a cut-off scale 
and we take $M_{\ast}=M_{Pl}$.
By this effective potential, the inflaton $\phi$ now rolls over. 
When $\phi = \phi_c \equiv \sqrt{\frac{N}{h}}m$, the scalar fields
$\chi, \,\bar{\chi}$ become tachyonic, and roll off to the vacuum with
non-vanishing VEVs of $\chi$  and $\bar{\chi}$.
Then, the inflationary process ends and 
the system approaches toward the ISS vacuum, which was said above.

In the above analysis, we have taken the global SUSY limit.
When we consider the supergravity scalar potential derived from 
the K\"ahler potential $K_{ISS}$ and the superpotential $W_{ISS}$,  
the mass term of $TrY = \sqrt{N}\phi$ has a 
correction of $\mathcal{O}(\mu^2/M_{Pl})$.
Such a correction can be negligible compared with the above 
1-loop correction when $\mu$ is sufficiently small.

In this model, the Hubble parameter during the inflation era 
is evaluated as 
\begin{equation}
H_{inf} = \frac{V^{1/2}}{\sqrt{3}M_{Pl}} \sim \frac{\sqrt{N}m^2}{\sqrt{3}M_{Pl}}.
\end{equation}
Furthermore, the superpotential during the inflation era 
is dominated by $Tr Y$, i.e.,
\begin{eqnarray}\label{eq:W-inf-ISS}
\langle W \rangle_{inf} \ \sim \ m^2 Tr Y.
\end{eqnarray}
These values are important as said in the previous section,  
and we will study them explicitly in the following section.
In this model, the sizes of the Hubble parameter $H_{inf}$ and 
the SUSY breaking are determined by the parameters, 
$m^2$ and $\mu^2$, respectively.
They are independent each other.
In addition, we are assuming $\mu^2 \ll m^2$ in Eq.~(\ref{eq:mu-m}).
Thus, we could realize the inflation with the low-energy 
SUSY breaking scenario, when the moduli are stabilized 
successfully.

Here, let us estimate the slow-roll parameters, $\epsilon$ and  
$\eta$.
We now parameterize the inflaton trajectory by $x \equiv \frac{\phi}{\phi_c}$, which leads to the effective potential of the form,
\begin{equation}
\begin{split}
V_{eff}(x) &= N m^4 \Biggl[ 1 + \frac{N h^2}{32\pi^2 } \biggl[ 2 \log{\left( \frac{hm^2 x^2}{M_{\ast}^2} \right)} \\
&\quad+ (x^2 + 1 )^2 \log{\left( 1 + \frac{1}{x^2} \right)} + (x^2 - 1 )^2 \log{\left( 1 - \frac{1}{x^2} \right)} \biggr]\Biggr].
\end{split}
\end{equation}
Then, the slow-roll parameters are given by
\begin{equation}
\begin{split}
&\epsilon = \frac{M_{Pl}^2}{2} \left( \frac{1}{V} \frac{\partial V}{\partial \phi_1} \right)^2 \simeq \frac{h^5 N M_{Pl}^2}{128 \pi^4 m^2} x^2 \left[ (x^2 - 1) \log{ \left( 1 - \frac{1}{x^2} \right)} + (x^2 + 1) \log{ \left( 1+\frac{1}{x^2} \right)} \right]^2,  \\
&\,\\ 
&\eta = M_{Pl}^2 \frac{1}{V} \frac{\partial^2 V}{\partial \phi_1^2} \simeq \frac{h^3 M_{Pl}^2}{8\pi^2 m^2} \left[ (3x^2-1)\log{ \left( 1 - \frac{1}{x^2} \right)} + (3x^2+1)\log{ \left( 1 + \frac{1}{x^2} \right)} \right].
\end{split}
\end{equation}
We assume that the inflation takes place in the $TrY$ direction, 
even when we include the modulus $T$. 
As shown in \cite{Nakai:2010km}, the number of e-foldings, $N_e \sim 54 \pm 7$ and the amplitude of curvature perturbation, $P_{\zeta}^{1/2} = 4.86 \times 10^{-5}$ are obtained as follow
\begin{equation}\label{eandp}
N_e \simeq 4\pi^2 \left( \frac{m}{M_{Pl}} \right)^2 \frac{x_e^2}{h^3}, \qquad
P_{\zeta}^{1/2} \simeq \frac{1}{\sqrt{2\epsilon}} \left( \frac{H_{inf}}{2\pi M_{Pl}} \right),
\end{equation}
where $x_e$ denotes $x_e=\phi_e/\phi_c$ with the initial value of 
the inflaton $\phi_e$.
Inserting the approximate expressions of these parameters, $H_{inf}=\sqrt{V/3M_{Pl}^2} \simeq \frac{\sqrt{N}m^2}{\sqrt{3}M_{Pl}} $ and $\epsilon \simeq \frac{h^2}{32\pi^2}\frac{N}{N_e}$, we can obtain
\begin{equation}
\frac{m}{M_{Pl}} \sim 10^{-3} \times h^{1/2},
\end{equation}
for $N_e \sim 54$.

\subsection{ Modulus stabilization and inflation}

Here, we consider the following K\"ahler potential 
and the superpotential,
\begin{eqnarray}\label{ISS-KKLT}
K = - \ln (T + T^*) +K_{ISS}, \qquad 
W= W_{ISS} -Ae^{-aT}.
\end{eqnarray}
First we study the ISS vacuum taking account the modulus $T$.
At the point, 
the dominant terms in the superpotential can be written as 
\begin{eqnarray}
W_{ISS} = w_0 - \mu^2\Phi_{aa} + \cdots .
\end{eqnarray}
At this minimum, the SUSY is broken by F-terms of $\Phi_{aa}$, i.e.
$|F^{\Phi}|=\mu^2$.
We assume that the VEV of $\Phi$ at the ISS vacuum is small.
Then, the VEV of $W_{ISS}$ is dominated by $w_0$.
The gravitino mass $m_{3/2}$ is evaluated as 
$m_{3/2} \simeq w_0/(T + T^*)^{3/2}$.
When we neglect $F^{\Phi}$, the modulus $T$ is stabilized at 
$D_TW=0$, where the modulus has the mass, 
\begin{eqnarray}
m_T \approx \frac{a}{\sqrt{T + T^*}} \ W_{ISS} \approx 
\frac{aw_0}{\sqrt{T + T^*}} .
\end{eqnarray}
Next, we take into account the effect of non-vanishing $F^{\Phi}$.
However, when $F^{\Phi}=\mu^2$ is smaller than the above 
modulus mass $m_T$, the modulus is stabilized almost around 
the point, $D_TW=0$.
Indeed, we fine-tune $w_0$ to lead almost vanishing vacuum energy, 
$V_0 \sim 0$ \cite{Abe:2006xp}, 
\begin{eqnarray}
w_0-Ae^{-aT} \simeq \frac{1}{\sqrt 3}\mu^2\left( 1 - \frac{2}{3} 
\left( \frac{\mu}{m_\Phi}\right)^2 \mu^2 \right),
\end{eqnarray}
where $m_\Phi$ is the mass of $\Phi$ at the ISS vacuum.
That is the F-term uplifting scenario \cite{Dudas:2006gr,Abe:2006xp}.
Thus, the modulus mass is heavier than the gravitino mass by 
the factor $a$.
The minimum of $T$ is slightly shifted from the point 
$D_TW=0$ and the non-vanishing $F^T$ appears as 
$F^T \simeq 3a^{-1}m_{3/2}$.
At any rate, we have $F^T \ll F^{\Phi} = \mu^2 $.
The gravitino mass at the ISS vacuum is evaluated as 
\begin{eqnarray}\label{eq:gravitino-mass}
m_{3/2} \simeq \frac{\mu^2}{\sqrt{3} (T + T^*)^{3/2}}
\left( 1 - \frac{2}{3} \left( \frac{\mu}{m_\Phi}\right)^2 \mu^2 \right).
\end{eqnarray}

Now, let us study the inflation with taking into account 
the modulus $T$.
During the inflation era, the first term $W_{ISS}$ in Eq.~(\ref{ISS-KKLT})
becomes large as Eq.~(\ref{eq:W-inf-ISS}).
Then, for a fixed value of $Tr Y$, the modulus $T$ is stabilized 
around $D_T W=0$, i.e.
\begin{eqnarray}\label{eq:DTW=0-3}
aT \sim  -log(m^2 Tr Y),
\end{eqnarray}
 as studied in Eqs.~(\ref{eq:DTW=0}) and
(\ref{eq:DTW=0-2})  of section 2.
Around that point, the modulus mass $m_T$ is estimated 
by Eq.~(\ref{eq:modulus-mass}) as 
\begin{eqnarray}
m_T  \approx  
\frac{a}{\sqrt{T + T^*}} \ W_{ISS} \approx 
\frac{am^2}{\sqrt{T + T^*}}  \ Tr  Y.
\end{eqnarray}
On the other hand, the Hubble parameter during the inflation era 
is evaluated as 
\begin{equation}\label{eq:Hubble}
H_{inf} = \frac{1}{\sqrt{3}}V^{1/2} \sim \frac{\sqrt{N}m^2}{\sqrt{3}(T
  + T^*)^{3/2}}.
\end{equation}
Thus, the modulus is not destabilized during the inflation era if 
\begin{eqnarray}\label{constrainT}
a(T+ T^*) \ Tr Y > \sqrt{N/3}.
\end{eqnarray}
For example, 
when $a(T+T^*)  \sim 50$ and $\sqrt{N/3} \sim 1$, 
this condition is satisfied for $Tr Y \geq {\cal O}(0.1) $.
Such a value of $T$, i.e. $a(T+ T^*)  \sim 50$ would be 
obtained for $m \sim 10^{14}$ GeV.
Then, a sufficiently large value of e-folding $N_e$ 
can be realized during $Tr Y \geq {\cal O}(0.1) $ 
when the initial condition of $\phi$ is large enough, 
e.g. $\phi_e=1$.
The magnitude of the Hubble parameter $H_{inf}$ during the 
inflation is determined by $m^2$ in Eq.~(\ref{eq:Hubble}), 
while the gravitino mass $m_{3/2}$ at the ISS vacuum is 
determined by $\mu^2$ in Eq.~(\ref{eq:gravitino-mass}).
Recall that $\mu^2$ and $m^2$ are independent each other 
and we are assuming $\mu^2 \ll m^2$ in Eq.~(\ref{eq:mu-m}).
Thus, we would realize much smaller value of $m_{3/2}$ 
than $H_{inf}$.

When we include the term $Ae^{-aT}$ in the superpotential 
and $-3\ln (T + T^*)$ in the K\"ahler potential, 
the supergravity scalar potential has a correction on 
the would-be inflaton mass and it is of 
${\cal O}(Ae^{-aT}/(T + T^*)^{3/2})$, which is 
estimated as 
\begin{eqnarray}
Ae^{-aT}/(T + T^*)^{3/2} \approx \frac{3 m^2 \ Tr Y}{a(T + \bar
  T)^{5/2}}.
\end{eqnarray}
Such a correction can be neglected for the inflation dynamics because 
$a = {\cal O}(10)$.

Figure 3 shows an example of the scalar potential $V$ of 
$Re T$ and $\phi$.
Figure 4 shows the same scalar potential $V$ of $Re T$
at the slice of $\phi =1.0$ (yellow curve), 
$\phi =0.8$ (red curve) and  
$\phi =0.3$ (green curve).
The inflation continues as far as the modulus is 
stabilized at the local minimum 
around $Re T \sim 1.6$, e.g. for 
$\phi =1.0$ (yellow curve) and $\phi =0.8$ (red curve) 
as well as smaller value of $\phi$.
(We will study later the point corresponding to 
$\phi =0.3$ (yellow curve).)
We may need fine-tuning to take the initial condition of 
$T$ as $Re T \sim 1.6$.

\begin{figure}[tbp]
 \begin{center}
  \begin{overpic}[width=10cm,clip]{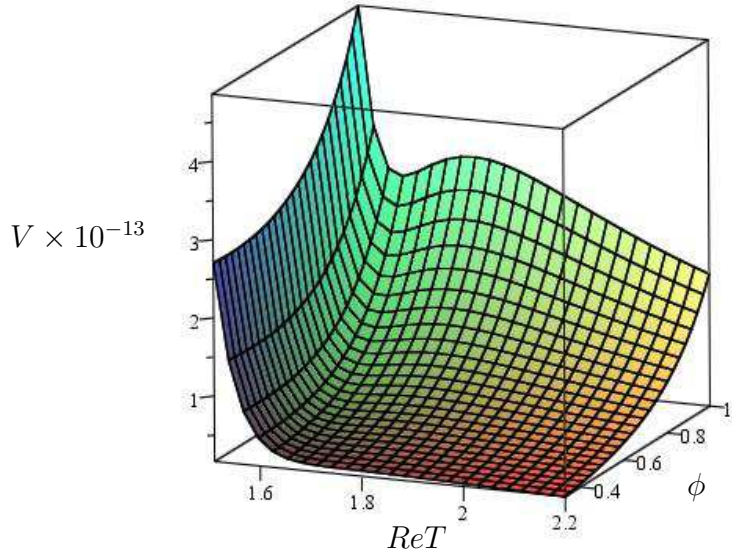}
   \put(-10,40){$V \times 10^{-13}$}
   \put(40,0){$Re T$}
   \put(80,7){$\phi$}
  \end{overpic}
\caption{The potential as a function of the moduli $Re T$ and the
  would-be inflaton $\phi$. Values of the parameters are taken as 
$A=1, a=10, w_0=-10^{-14}, N=3, m=7.0\times 10^{-4}, h=0.5$. }
 \end{center}
\end{figure}

\begin{figure}[tbp]
 \begin{center}
  \begin{overpic}[width=7cm,clip]{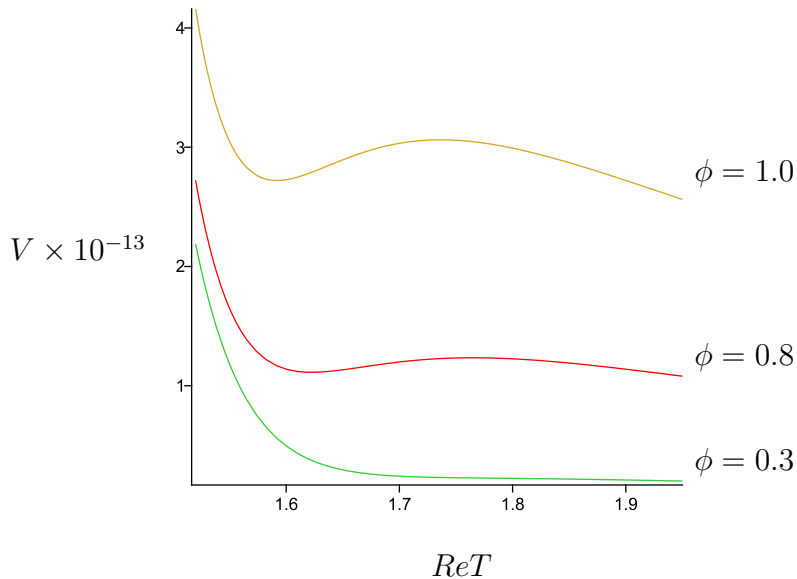}
   \put(-30,50){$V \times 10^{-13}$}
   \put(50,-10){$Re T$}
   \put(100,65){$\phi=1.0$}
   \put(100,30){$\phi=0.8$}
   \put(100,10){$\phi=0.3$}
  \end{overpic}
\vspace{4mm}
\caption{The same scalar potential as Figure 3.
The top (yellow) curve corresponds to the slice of $\phi =1.0$.
Similarly, the second (red) curve and lowest (green) curve 
correspond to   $\phi =0.8$ and $0.3$, respectively.}
 \end{center}
\end{figure}

We study the final stage of the inflation.
At least, the waterfall point, $\phi_c = \sqrt{\frac{N}{h}}m$ should
be larger than the point of $\phi$ leading to $m_T \sim H_{inf}$, 
since the moduli is destabilized at $m_T \sim H_{inf}$.
We denote a value of the would-be inflaton by 
$\phi_\ast( \equiv g M_{Pl},g<1)$, where the 
modulus direction becomes tachyonic.
At $\phi=\phi_*$, we would have 
\begin{eqnarray}
m_T \approx \frac{am^2}{\sqrt{T + T^*}}  \ Tr  Y \sim H_{inf}.
\end{eqnarray}
Then, the condition that the modulus is not destabilized during the
whole inflation process is written by 
\begin{equation}\label{eq:inflation-condition}
\phi_\ast < \phi_c < \phi_e.
\end{equation}
We assume that the initial value of the would-be inflaton is $M_{Pl}$, 
i.e. $\phi_e=M_{Pl}$. 
Then, for $N_e \sim 54$, we obtain $Nh^2 = \frac{4\pi^2}{N_e} \sim
0.73$ and $\phi_e=10^{-3}\sqrt{N}M_{Pl}$, where we used Eq. (\ref{eandp}). 
The above condition (\ref{eq:inflation-condition}) becomes
\begin{equation}
10^6 g^2 < N < 10^6 \quad \text{or} \quad 10^{-3} < h < 10^{-3}g^{-1}.
\end{equation}
In our model, a value of $g$ is of $\mathcal{O}(0.1)$, 
then we need a large number of $N\sim\mathcal{O}(10^5)$.

At any rate, when the above situation is satisfied, 
the waterfall occurs at $\phi = \phi_c$ by the tachyonic masses 
of $\chi$ and $\bar{\chi}$ directions.
During such a waterfall the $\chi$ and $\bar \chi$ develop 
their values, which we parameterize them as 
$\chi \bar{\chi} = \alpha m^2/h$, where $\alpha$ varies $\alpha=0 -1$.
Then, the values of superpotential and Hubble 
parameter vary like $W= \sqrt{N}m^2 \phi(\alpha-1)$ and
$H=\sqrt{N}m^2(1 - \alpha)$. Thus, the condition that the moduli is
stabilized during the waterfall is the same as
Eq. (\ref{constrainT}). 
During the waterfall, however, $\phi$ rolls to the ISS point which is
smaller than $\phi_\ast$. 
Thus Eq. (\ref{constrainT}) would not be satisfied during the
waterfall.
The moduli would not be stabilized during the waterfall.

\subsection{Waterfall by modulus}

The inflation scenario in the previous section can be 
realized if the relation $\phi_* < \phi_c$ is satisfied.
That requires that $N = {\cal O}(10^5)$.
That might be too large to construct explicit (string) models.
If the value of $N$ is small like $N={\cal O}(1-10)$ and 
the situation $\phi_* > \phi_c$ is realized, 
the modulus is destabilized before the waterfall by tachyonic masses 
of $\chi$ and $\bar \chi$.
That corresponds to the lowest (green) curve in Figure 4. 
Then, the inflation ends.
The inflation itself has no problem.
That is, we can realize sufficiently large e-folding 
before the destabilization of the modulus $T$.
However, once the modulus $T$ is destabilized with 
the superpotential (\ref{ISS-KKLT}), 
the modulus $T$ runs away to infinity.

We can avoid such a runaway behavior by adding 
the non-perturbative term $Be^{bT}$ with the positive exponent $b$,
i.e.,
\begin{eqnarray}\label{eq:W-water}
W= W_{ISS} -Ae^{-aT}+Be^{bT}.
\end{eqnarray}
As said above, the previous scenario with $B=0$ has no problem 
for the inflation to realize a sufficiently large value of e-folding.
Hence, we take values of $B$ and $b$ such that they are small enough not to 
affect the inflation around $T_{inf}$, which denotes 
the value of $T$ determined by $D_TW=0$ with $B=0$, i.e. 
Eq.~(\ref{eq:DTW=0-3}).

In this case, the inflation terminates due to the instability 
caused by the moduli $T$, but not $\chi,\bar \chi$ fields.
That is, the modulus becomes the waterfall field.
That has new and interesting effects on several cosmological aspects.
However, studies on such aspects are beyond the purpose of this paper.
We would study them elsewhere.
Figure 5 shows the scalar potential $V$ of 
$Re T$ and $\phi$ for the superpotential (\ref{eq:W-water}).


\begin{figure}[tbp]
\begin{center}
\begin{overpic}[width=10cm,clip]{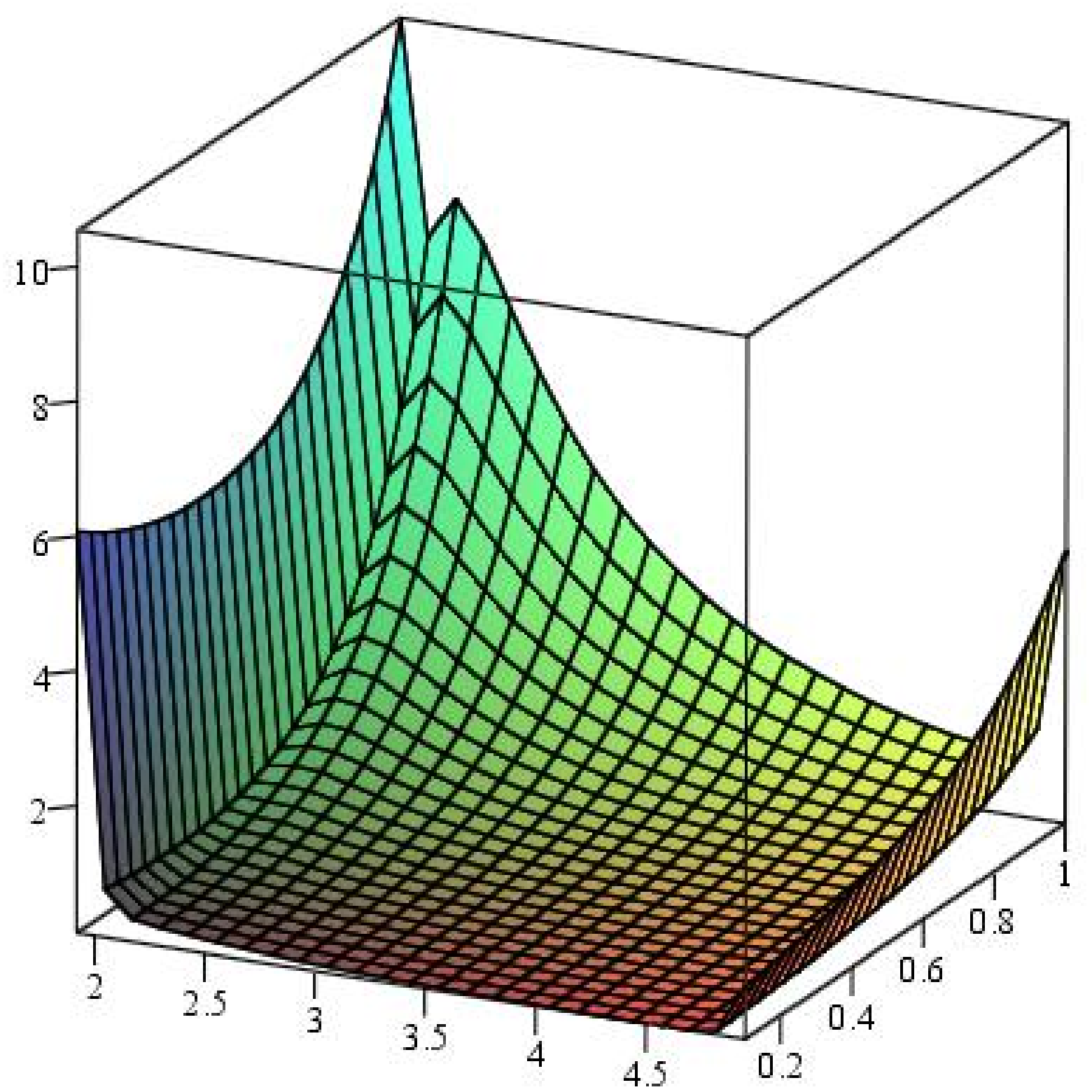}
\put(-10,50){$V \times 10^{-17}$}
\put(80,10){$\phi$}
\put(35,5){$Re T$}
\end{overpic}

\begin{overpic}[width=9cm,clip]{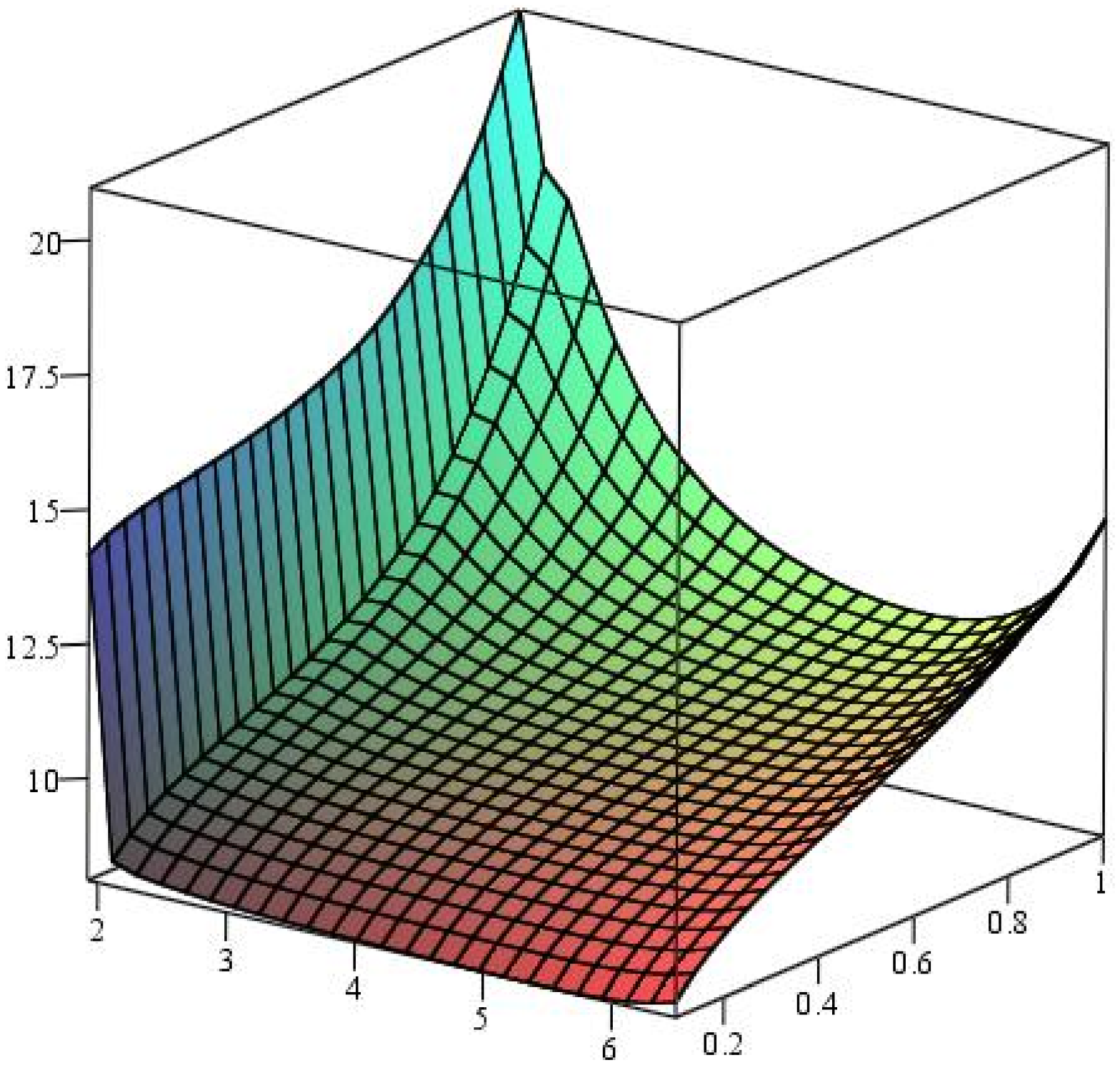}
\put(-20,50){$V \times 10^{-17}$}
\put(79,5){$\phi$}
\put(30,1){$Re T$}
\end{overpic}
\caption{The potential including positive exponent term. We take the
  values of parameters as $b=10,B=10^{-30}$ in the upper figure
  and $b=1,B=10^{-10}$ in the lower figure. The other parameters
  are taken as the same values as those in Figures 3.}
\end{center}
\end{figure}

At any rate, the runaway of $T$ to infinity is avoided by the term with 
the positive exponent, $Be^{bT}$.
Then, the field $\phi$ rolls down around $\phi_c$, where 
$\chi, \bar \chi$ fields become tachyonic.
Then, the system approaches toward the ISS vacuum.
At that point the dominant superpotential can be written by 
\begin{eqnarray}
W = w_0 - \mu^2\Phi_{aa} + Be^{bT} + \cdots .
\end{eqnarray}
Similar to the discussion at the beginning of section 3.2, 
the modulus $T$ is stabilized around $D_TW=0$, where 
the modulus has the mass, 
\begin{eqnarray}
m_T \approx
\frac{bw_0}{\sqrt{T + T^*}} .
\end{eqnarray}
The SUSY breaking is dominated by $F^\Phi = \mu^2$, while 
the $F^T$ is suppressed as $F^T \simeq 3b^{-1}m_{3/2} \ll F^\Phi =
\mu^2 \approx m_{3/2}$.
Recall again that $\mu^2$ and $m^2$ are independent each other.
Thus, we would realize much smaller value of $m_{3/2}$ 
than $H_{inf}$.

\subsection{Positive exponent}

It would be simple to consider the following superpotential 
only including $W_{ISS}$ and a single modulus-dependent 
term with a positive exponent, i.e.,
\begin{eqnarray}
W=W_{ISS} + Ae^{aT},
\end{eqnarray}
where $A \ll 1$ and $a >0$.
During the inflation era, the dominant terms in the superpotential 
are written by 
\begin{eqnarray}\label{eq:ISS-inf-positive}
W = - m^2 Tr Y + Ae^{aT} + \cdots .
\end{eqnarray}
We take the initial condition $\phi_e =1$.
Then, the inflation occurs.
As far as $\phi$ is large enough, 
the modulus $T$ is stabilized at $D_T W=0$ for the above 
superpotential (\ref{eq:ISS-inf-positive}), i.e. 
\begin{eqnarray}
aT \sim  -log(m^2 Tr Y), 
\end{eqnarray}
and the modulus $T$ has the mass $m_T \simeq am^2 Tr Y/\sqrt(T + \bar
T)$.
Thus, a sufficiently large value of e-folding can 
be realized, similar to that in the previous sections.
As the field $\phi$ rolls down the scalar potential and 
its value becomes of ${\cal O}(0.1)$, the modulus 
mass becomes of ${\cal O}(H_{inf})$.
The modulus shifts its value, but the modulus 
is not destabilized, because there are quite high 
potential barriers at both small and large values of 
$Re T$.
Figure 6 shows the scalar potential of $Re T$ at the 
slice of $\phi= 1.0, 0.7$ and $0.1$.
Then, at $\phi = \phi_c$, the tachyonic direction appears 
along $\chi$ and $\bar \chi$, but not $T$.
The system approaches toward the ISS vacuum, where
the dominant terms in the superpotential can be written by
\begin{eqnarray}
W = w_0 - \mu^2\Phi_{aa} + Ae^{aT} + \cdots .
\end{eqnarray}
At the ISS vacuum, the large modulus mass is realized 
and we can realize the gravitino mass $m_{3/2}$ 
smaller than $H_{inf}$.

\begin{figure}
\begin{center}
\begin{overpic}[width=7cm,clip]{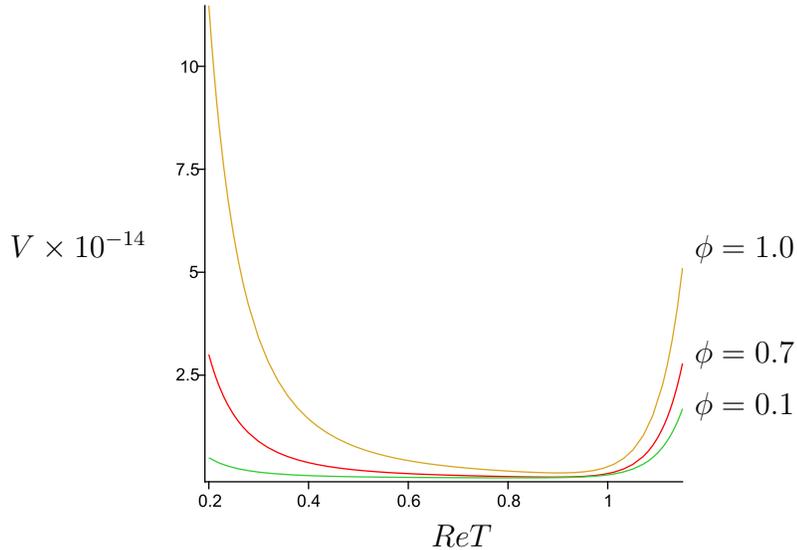}
\put(-30,50){$V \times 10^{-14}$}
\put(50,-5){$Re T$}
\put(100,50){$\phi=1.0$}
\put(100,30){$\phi=0.7$}
\put(100,20){$\phi=0.1$}
\end{overpic}
\caption{The potential of the positive exponent superpotential. We
  take the values of parameters as
  $w_0=-10^{-14},A=-10^{-12},a=-10,m=10^{-4},N=3$. 
The top (yellow), middle (red) and lowest (green) curves show 
that when $\phi=1.0,\phi=0.7,\phi=0.1$.}
\end{center}
\end{figure}

\section{Conclusion}

We have studied the inflation scenario from the viewpoint 
of the moduli stabilization.
We have proposed the scenario that the superpotential 
has a large value during the inflation era and it 
induces a large mass of the modulus field.
Then, we could realize the inflational expansion of the 
Universe with a sufficiently large value of e-folding $N_e$ 
without destabilizing the modulus.
We study our scenario by using a simple model, 
the deformed ISS inflation model.
It is found that in that model a sufficiently large e-folding 
can be obtained during the inflation, but 
at the final stage the modulus may be destabilized.
Then, the modulus would run away to infinity 
in the simple KKLT superpotential.
Thus,  we would need another bump/barrier to 
avoid the runaway behavior of the modulus.
Indeed, we have added the term with a positive exponent, 
although another type of potential bump may also be useful.
As a result, this model shows a new aspect.
The inflation is terminated when the modulus mass becomes 
tachyonic.
That is, the modulus becomes the waterfall field.
That has new and interesting effects on several cosmological aspects.
We would study them elsewhere.
At any rate, the modulus is not destabilized by the term 
with a positive exponent.
Then, the system approaches toward the ISS vacuum.
At the true (ISS) vacuum with the modulus stabilized, 
we can obtain the gravitino mass $m_{3/2}$, 
which is much smaller than the Hubble parameter 
$H_{inf}$ during the inflation.

On the other hand, other inflation models would 
lead to a different behavior.
The modulus might be stabilized in the whole inflation 
process in another inflation model with a large 
value of superpotential.
It would be interesting to apply our studies 
to several types of inflation models.

\subsection*{Acknowledgement}
The authors would like to thank Y.~Nakai and O.~Seto 
for useful discussions.
T.~K. is supported in part by the Grant-in-Aid for 
Scientific Research No.~20540266 and 
the Grant-in-Aid for the Global COE 
Program "The Next Generation of Physics, Spun from Universality and 
Emergence" from the Ministry of Education, Culture,Sports, Science and 
Technology of Japan.

\end{document}